\RequirePackage{ifpdf}
\ifpdf 
\documentclass[pdftex]{sigma}
\else
\documentclass{sigma}
\fi

\numberwithin{equation}{section}

\begin{document}

\allowdisplaybreaks

\renewcommand{\PaperNumber}{092}

\FirstPageHeading

\ShortArticleName{On Non-Point Invertible Transformations}

\ArticleName{On Non-Point Invertible Transformations\\ of Dif\/ference and Dif\/ferential-Dif\/ference Equations}

\Author{Sergey Ya. STARTSEV}

\AuthorNameForHeading{S.Ya. Startsev}
\Address{Ufa Institute of Mathematics, Russian Academy of Sciences,\\ 112 Chernyshevsky Str., Ufa, 450077, Russia}
\Email{\href{mailto:startsev@anrb.ru}{startsev@anrb.ru}}

\ArticleDates{Received October 04, 2010, in f\/inal form December 03, 2010;  Published online December 11, 2010}

\Abstract{Non-point invertible transformations are completely described for dif\/ference equations on the quad-graph and for their dif\/ferential-dif\/ference analogues. As an illustration, these transformations are used to construct new examples of integrable equations and autotransformations of the Hietarinta equation.}


\Classification{39A14; 34K99; 37K35}

\section{Introduction}

The present paper is devoted to invertible transformations for both discrete equations of the form
\[u_{i+1,j+1}=F(u_{i,j},u_{i+1,j},u_{i,j+1}),\]
and ``semi-discrete'' chains of the dif\/ferential equations
\[(u_{i+1})_x=F(x,u_i,u_{i+1},(u_{i})_x).\]
Here $i$ and $j$ are integers, $x$ is a continuous variable, $u$ is a function of $i$, $j$ and $i$, $x$ for the f\/irst and the second equation, respectively.
From now on, we shall omit $i$ and $j$ for brevity and, in particular, write the above equations in the form
\begin{gather}\label{uij}
u_{1,1}=F(u,u_{1,0},u_{0,1})
\end{gather}
and
\begin{gather}\label{uix}
(u_1)_x=F(x,u,u_1,u_x).
\end{gather}
We assume that $F_u F_{u_{1,0}} F_{u_{1,0}} \ne 0$ for equation~\eqref{uij} and $F_{u_{x}} \ne 0$ for equation~\eqref{uix}. These conditions allows us to rewrite equation~\eqref{uij} in any of the following forms
\begin{gather}\label{umm}
u_{-1,-1}=\overline{F}(u,u_{-1,0},u_{0,-1}),
\\
\label{upm}
u_{1,-1}=\hat{F}(u,u_{1,0},u_{0,-1}),
\\
\label{ump}
u_{-1,1}=\tilde{F}(u,u_{-1,0},u_{0,1}),
\end{gather}
and equation~\eqref{uix} -- in the form
\begin{gather}\label{umx}
(u_{-1})_x=\tilde{F}(x,u,u_{-1},u_x).
\end{gather}
Therefore, all ``mixed shifts'' $u_{m,n}:=u_{i+m,j+n}$ (for both positive and negative non-zero~$n$ and~$m$) can be expressed in terms of {\it dynamical variables} $u_{k,0}$, $u_{0,l}$ by virtue of equations~\eqref{uij}, \eqref{umm}--\eqref{ump}. (A more detailed explanation of the dynamical variables, the notation $u_{m,n}$ and the recursive procedure of the mixed shift elimination can be found, for example, in~\cite{Mikh,LY}.) Analogously, $u_m^{(n)}:=\partial ^n u_{i+m} /\partial x^n$ for any non-zero $m \in \mathbb Z$ and $n \in \mathbb N$ can be expressed in terms of $x$ and dynamical variables $u_l:=u_{i+l}$, $u^{(k)}:=\partial ^k u_i /\partial x^k$ by virtue of equations~\eqref{uix}, \eqref{umx}. The notation $g[u]$ means that the function $g$ depends on a f\/inite number of the dynamical variables (and $x$ if we consider equation~\eqref{uix}). The considerations in this paper are local (for example, we use the local implicit function theorem to obtain \eqref{umm}--\eqref{umx}) and, for simplicity, all functions are assumed to be locally analytical.

In addition to the point transformations $v=g(u)$, some of the equations~\eqref{uij} and~\eqref{uix} admit non-point transformations $v=g[u]$ which are invertible in the sense of \cite{SokSv}. For example, the dif\/ferential substitutions
\begin{gather}\label{sg2csg}
v=\frac{u_x - \sin u}{2}
\end{gather}
maps solutions of the dif\/ferential-dif\/ference sine-Gordon equation \cite{Hir77,Orf}
\begin{gather}\label{sdsg}
(u_1)_x - \sin u_1= u_x +\sin u
\end{gather}
into solutions of the equation
\begin{gather}\label{sdcsg}
\frac{(v_1-v)_x}{\sqrt{1-(v_1-v)^2}}=\pm (v_1+v) ,
\end{gather}
which is a semi-discrete version of the complex sine-Gordon equation. Here the sign of the right-hand side of equation~\eqref{sdcsg} coincides with the sign of the $\cos u$ value\footnote{A local transformation of an equation may, generally speaking, generate dif\/ferent equations for dif\/ferent domains of the ``jet space''. This is true for both point and non-point local transformations.}. Indeed, $v_1=(u_x+ \sin u)/2$ follows from equation~\eqref{sdsg} and, together with \eqref{sg2csg}, gives us
\begin{gather*}
 u_x=v_1+v,\quad \sin u= v_1-v \quad \Longrightarrow \quad
(v_1-v)_x= u_x \cos u =\pm (v_1+v) \sqrt{1-(v_1-v)^2}.
\end{gather*}
The inverse transformation can be found in \cite{NQC83}: the formula
$u=\frac{\pi}{2} \pm (\arcsin (v_1-v)-\frac{\pi}{2})$ maps any real solution of equation~\eqref{sdcsg} into a solution of equation~\eqref{sdsg}.

This example belongs to the following class of non-point invertible transformations introduced in \cite{Yam90}. Let functions $\varphi (x,y,z)$, $\psi (x,y,z)$ satisfy the condition $\varphi _y \psi _z - \varphi _z \psi _y \ne 0$ and equation~\eqref{uix} can be written in the form
\begin{gather}\label{ab}
\varphi (x,u_1,(u_1)_x)=\psi (x,u,u_x).
\end{gather}
Then we rewrite \eqref{ab} in the form of the system
\begin{gather}\label{abs}
v=\varphi (x,u,u_x), \qquad v_1=\psi (x,u,u_x),
\end{gather}
express $u$, $u_x$ in terms of $v$, $v_1$ from \eqref{abs} and obtain
\begin{gather}\label{pqs}
u=p(x,v,v_1), \qquad u_x=q(x,v,v_1).
\end{gather}
The system \eqref{pqs} is equivalent to the equation
\begin{gather}\label{pq}
D_x(p(x,v,v_1))=q(x,v,v_1),
\end{gather}
where $D_x$ denotes the total derivative with respect to $x$. The substitution $v=\varphi (x,u,u_x)$ maps solutions of \eqref{ab} into solutions of \eqref{pq} and the transformation $u=p(x,v,v_1)$ maps
solutions of \eqref{pq} back into solutions of \eqref{ab}.

It is easy to see that the same scheme can be applied to the pure discrete equations of the form
\begin{gather}\label{abd}
\varphi (u_{0,1},u_{1,1})=\psi (u,u_{1,0}),
\end{gather}
where $\varphi (y,z)$ and $\psi (y,z)$ are functionally independent. Indeed, expressing $u$ and $u_{1,0}$ from
\begin{gather}\label{absd}
v=\varphi (u,u_{1,0}),\qquad v_{0,1}=\psi (u,u_{1,0}),
\end{gather}
we obtain
\begin{gather}\label{pqsd}
u=p(v,v_{0,1}),\qquad u_{1,0}=q(v,v_{0,1})
\end{gather}
and rewrite \eqref{pqsd} in the form of the equivalent equation
\begin{gather}\label{pqd}
p(v_{1,0},v_{1,1})=q(v,v_{0,1}).
\end{gather}
Thus, the transformation $v=\varphi (u,u_{1,0})$ maps solutions of \eqref{abd} into solutions of \eqref{pqd} and the inverse transformation $u=p(v,v_{0,1})$ maps solutions of \eqref{pqd} back into solutions of \eqref{abd}. The transformations \eqref{abd}--\eqref{pqd} were, in fact, used in \cite{Yam94} without explicit formulation of the above scheme.

The main result of this paper is the proof of the following statement: any invertible transformation of equations~\eqref{uij}, \eqref{uix} is a composition of shifts, point transformations and transformations \eqref{ab}--\eqref{pq}, \eqref{abd}--\eqref{pqd}. Roughly speaking, equations~\eqref{uij} and \eqref{uix} have no non-point invertible transformations other than \eqref{abd}--\eqref{pqd} and \eqref{ab}--\eqref{pq}, respectively. The proof is similar to that was used in~\cite{SokSv} for continuous equations (hyperbolic PDEs).

The invertible transformations allow us to obtain objects associated with integrability of  equations~\eqref{pq}, \eqref{pqd} (such as conservation laws and higher symmetries) from the correspon\-ding objects of equations~\eqref{ab}, \eqref{abd} because we can express shifts and derivatives of $u$ in terms of shifts and derivatives of $v$. Therefore, the invertible transformations may be useful for constructing new examples of integrable equations of the form~\eqref{uij}, \eqref{uix}. To illustrate this, in Section~\ref{s4} we construct Darboux integrable equations related via invertible transformations to dif\/ference and dif\/ferential-dif\/ference analogues of the Liouville equation. In addition, an example of constructing an equation possessing the higher symmetries is contained at the end of Section~\ref{s2}. In this section we also demonstrate that the scheme \eqref{abd}--\eqref{pqd} generates autotransformations of the Hietarinta equation.

\section{Invertible transformations of discrete equations}\label{s2}

We let $T_i$ and $T_j$ denote the operators of the forward shifts in $i$ and $j$ by virtue of equation~\eqref{uij}. These operators are def\/ined by the following rules:
$T_i(f(a,b,c,\dots))=f(T_i(a),T_i(b),T_i(c),\dots)$ and $T_j(f(a,b,c,\dots))=f(T_j(a),T_j(b),T_j(c),\dots)$ for any function $f$;
$T_i(u_{m,0})=u_{m+1,0}$ and $T_j(u_{0,n})=u_{0,n+1}$;
$T_i(u_{0,n})=T_j^{n-1}(F)$ for positive $n$ and $T_i(u_{0,n})=T_j^{n+1}(\hat{F})$ for negative $n$, $T_j(u_{m,0})=T_i^{m-1}(F)$ for positive $m$ and $T_j(u_{m,0})=T_i^{m+1}(\tilde{F})$ for negative $m$ (i.e.\ mixed variables $u_{1,n}$ and $u_{m,1}$ are expressed in terms of the dynamical variables by virtue of equations~\eqref{uij}, \eqref{upm}, \eqref{ump}).
The inverse (backward) shift operators $T_i^{-1}$ and $T_j^{-1}$ are def\/ined in the similar way.

\begin{definition}\label{dtran}
We say that a transformation $v=f[u]$ maps the equation~\eqref{uij} into an equation $v_{1,1}=G(v,v_{1,0},v_{0,1})$ if
\begin{gather}\label{dtr}
T_i T_j (f)=G(f,T_i(f),T_j(f)).
\end{gather}
\end{definition}

\begin{definition}\label{it}
A transformation $v=f[u]$ of equation~\eqref{uij} is called invertible if any of the dynamical variables $u$, $u_{k,0}$, $u_{0,l}$, $k,l \in \mathbb Z$, can be expressed as a function of a f\/inite subset of the variables
\begin{gather}\label{vvar}
v := f,\qquad v_{r,0} := T_i^{r}(f),\qquad v_{0,s}:=T_j^s(f),\qquad r,s \in \mathbb Z.
\end{gather}
\end{definition}
We exclude all mixed variables $v_{r,s}$, $rs \ne 0$, from \eqref{vvar} because we consider only the cases when the transformation maps \eqref{uij} into an equation of the form
\begin{gather}\label{vij}
v_{1,1}=G(v,v_{1,0},v_{0,1}), \qquad G_v G_{v_{1,0}} G_{v_{0,1}} \ne 0
\end{gather}
and the mixed variables can be expressed in terms of \eqref{vvar} by virtue of this equation.

It is easy to see that any shift $w=v_{r,s}$ maps equation~\eqref{vij} into equation~\eqref{vij} again and the composition of the shift and an invertible transformation $v=f[u]$ is invertible too. This leads to the following
\begin{definition}\label{equiv}
Transformations $v=f[u]$ and $w=g[u]$ are called equivalent if there exist $r,s \in \mathbb Z$ such that $g= T_i^s T_j^r(f)$.
\end{definition}
\begin{proposition}\label{char}
Let $v=g[u]$ be an invertible transformation that maps equation~\eqref{uij} into equation~\eqref{vij}. Then this transformation is equivalent to either a transformation of the form
\begin{gather}\label{itr}
w=f(u, u_{1,0}, u_{2,0}, \dots, u_{m,0}),
\end{gather}
or a transformation of the form
\begin{gather}\label{jtr}
w=f(u, u_{0,1}, u_{0,2}, \dots, u_{0,n}).
\end{gather}
\end{proposition}
\begin{proof}
The transformation is equivalent to that of the form
\begin{gather}\label{u2v}
v=h(u,u_{1,0}, \dots,u_{k,0}, u_{0,1}, \dots, u_{0,l})
\end{gather}
because we can eliminate ``negative'' variables $u_{r,0}$, $u_{0,s}$, $r,s < 0$ from the transformation by shifts of $g$.
We can express $u$ as
\begin{gather}
u=P(v_{a,0}, v_{a+1,0}, \dots, v_{b,0}, v_{0,c}, v_{0,c+1}, \dots, v_{0,d}) \nonumber \\
\phantom{u} := P\big(T_i^a(h), T_i^{a+1}(h), \dots, T_i^b(h), T_j^c(h), T_j^{c+1}(h), \dots, T_j^d(h)\big)\label{v2u}
\end{gather}
if the transformation is invertible. Dif\/ferentiating equation~\eqref{v2u} with respect to $u_{k+b,0}$, we obtain $P_{v_{b,0}} T_i^b(h_{u_{k,0}})=0\: \Rightarrow\: P_{v_{b,0}}=0 $ if $b, k > 0$. The analogous reasoning gives $P_{v_{0,d}}=0$ if $d, l > 0$. Thus, $b, d \le 0$ if $k l \ne 0$.

Let $\left( T_i^{-1}(h)\right) _{u_{-1,0}} \big( T_j^{-1}(h)\big) _{u_{0,-1}} \ne 0$. Then $\left( T_i^{a}(h)\right) _{u_{a,0}} \big( T_j^{c}(h)\big) _{u_{0,c}} \ne 0$ for any negative $a$ and $c$, and we obtain $P_{v_{a,0}}=P_{v_{0,c}}=0$ by dif\/ferentiating equation~\eqref{v2u} with respect to $u_{a,0}$ and $u_{0,c}$. Therefore, either $a=c=b=d=0$ (i.e.~$u=P(h)$ that is possible only if $k=l=0$) or $\left( T_i^{-1}(h)\right) _{u_{-1,0}} \big( T_j^{-1}(h)\big) _{u_{0,-1}} = 0$. The latter equality means that either $T_i^{-1}(h) = \tilde{h}(u, u_{1,0}, \dots, u_{k-1,0}, u_{0,1}, \dots, u_{0,l})$ or $T_j^{-1}(h) = \tilde{h}(u, u_{1,0}, \dots, u_{k,0}, u_{0,1}, \dots, u_{0,l-1})$, i.e.\ any invertible transformation of the form~\eqref{u2v} with $k l \ne 0$ is equivalent to a transformation $\tilde{v}=\tilde{h}(u, u_{1,0}, \dots, u_{\tilde{k},0}, u_{0,1}, \dots, u_{0,\tilde{l}})$ such that $\tilde{k} \tilde{l} < k l$. Applying this conclusion several times, we obtain that \eqref{u2v} is equivalent to a transformation $w= f(u, u_{1,0}, \dots, u_{m,0}, u_{0,1}, \dots, u_{0,n})$ with $m n = 0$. \end{proof}

\begin{definition}
A transformation is called non-point if this transformation is not equivalent to any point transformation of the form $w=g(u)$.
\end{definition}
Because the transformations $v=f(u_{m,0})$ and $v=f(u_{0,n})$ are equivalent to the point transformation $w=f(u)$, a transformation of the form~\eqref{itr} or~\eqref{jtr} is non-point only if $f$ depends on more than one variable. We use only this property of the non-point transformations in the proof of the following

\begin{theorem}\label{form}
Let a non-point invertible transformation of the form~\eqref{itr}
map equation~\eqref{uij} into equation~\eqref{vij}. Then equation~\eqref{uij} can be written in the form $\varphi (u_{0,1},u_{1,1})=\psi (u,u_{1,0})$, where $\varphi (y,z)$ and $\psi (y,z)$ are functionally independent, and the transformation is equivalent to the composition of the invertible transformation $w=\varphi (u,u_{1,0})$ and an invertible transformation of the form $v=h(w, w_{1,0}, w_{2,0}, \dots, w_{m-1,0})$. In particular, any non-point invertible transformation of the form $v=f(u, u_{1,0})$ is equivalent to the composition of the transformation $w=\varphi (u,u_{1,0})$ and a point transformation $v=h(w)$.
\end{theorem}

\begin{proof} If $f_{u}=0$ and $s$ is the smallest integer for which $f_{u_{s,0}} \ne 0$, then the equivalent transformation $v=T_i^{-s}(f[u])$ depends on $u$. Therefore, we can, without loss of generality, assume that $f_u \ne 0$. We also can write
\begin{gather*}
u_{l,0}=P_l(v_{a,0}, v_{a+1,0}, \dots, v_{b,0}, v_{0,c}, v_{0,c+1}, \dots, v_{0,d})
\\
\phantom{u_{l,0}}{} = P_l\big(T_i^a(f), T_i^{a+1}(f), \dots, T_i^b(f), T_j^c(f), T_j^{c+1}(f), \dots, T_j^d(f)\big),\qquad l=\overline{0,m}
\end{gather*}
because the transformation is invertible. Here the notation $l=\overline{0,m}$ means that $l$ runs over all integers from $0$ to $m$. Dif\/ferentiating these equalities with respect to $u_{a,0}$, we obtain $(P_l)_{v_{a,0}} T_i^a(f_u)=0  \Rightarrow  (P_l)_{v_{a,0}}=0$ if $a < 0$. The similar reasoning gives $(P_l)_{v_{b,0}}=0$ if $b > 0$. Thus,
\[u_{l,0}=P_l\big(T_j^c(f), T_j^{c+1}(f), \dots, T_j^d(f)\big),\qquad l=\overline{0,m}.\]

Let $c < 0$ and $s$ be the biggest negative integer such that $(T_j^{s}(f))_{u_{0,-1}} \ne 0$. If $s \ge c$, then $T_j^{c}(f)$ depends on $u_{0,c-s-1}$ and $(P_l)_{u_{0,c-s-1}}= (P_l)_{v_{0,c}} (T_j^{c}(f))_{u_{0,c-s-1}}=0\: \Rightarrow\: (P_l)_{v_{0,c}}=0$. Hence $s < c$, i.e. $(T_j^{r}(f))_{u_{0,-1}} = 0$ for all $r \ge c$. This implies $T_j^{c}(f)= g(u, u_{1,0},\dots, u_{m,0})$ and
\begin{gather}\label{pl}
u_{l,0}=P_l\big(g, T_j(g), \dots, T_j^{\tilde{d}}(g)\big),\qquad l=\overline{0,m}.
\end{gather}
If $c \ge 0$, then equations~\eqref{pl} holds too, with $g=f$ and $\tilde{d}=d$.

Repeating the above argumentation, we prove that $(T_j^r(g))_{u_{0,1}}=0$ for all $r \le \tilde{d}$. Let us consider the operators $X=T_j^{-1} \partial _{u_{0,1}} T_j$ (cf.~\cite{Hab}) and $Y=[\partial _{u_{0,-1}},X]$, where $\partial _z := \frac{\partial}{\partial z}$. These operators have the form
\[ X = \partial _u + \sum_{l=1}^{m} \xi _l \partial _{u_{l,0}},\qquad
Y = \sum_{l=1}^{m} \nu _l \partial _{u_{l,0}}\]
for functions of $u,u_{1,0},\dots,u_{m,0}$. According to equation~\eqref{pl}, the set $\{g, T_j(g), \dots, T_j^{\tilde{d}}(g)\}$ must contain $m+1$ functionally independent functions because $u_{l,0}$, $l=\overline{0,m}$, are functionally independent. Hence the system $X(z)=0$, $Y(z)=0$ has $m$ functionally independent solutions depending on $u, u_{1,0}, \dots, u_{m,0}$ and the vectors $(1,\xi _1, \dots, \xi _m)$, $(0,\nu _1, \dots, \nu _m)$ must be collinear. The latter is possible only if $\nu _l = 0$ for all $l=\overline{0,m}$. In particular,
\begin{gather*}
\nu _1 = [T_j^{-1}(F_{u_{0,1}})]_{u_{0,-1}} = 0\  \Rightarrow\  T_j^{-1}(F_{u_{0,1}})=\alpha (u,u_{1,0}) \  \Rightarrow\  F_{u_{0,1}}=T_j(\alpha) =\alpha (u_{0,1}, F)  \\
\qquad \Rightarrow\  F_{u u_{0,1}} = \alpha _{u_{1,0}} (u_{0,1}, F) F_u,\quad  F_{u_{1,0} u_{0,1}} = \alpha _{u_{1,0}} (u_{0,1}, F) F_{u_{1,0}}\\
\qquad \Rightarrow\  (\ln (F_{u_{1,0}}) - \ln (F_{u}) )_{u_{0,1}} =0 \  \Rightarrow
\ F_{u_{1,0}} - \beta (u,u_{1,0})F_{u} =0 \\
\qquad  \Rightarrow\  F=E(\psi (u,u_{1,0}), u_{0,1}),
\end{gather*}
where $\psi$ is a solution of the equation $\psi _{u_{1,0}} - \beta (u,u_{1,0}) \psi _u = 0$. Thus, equation~\eqref{uij} can be written in the form~\eqref{abd}.

We can express $g$ in terms of $u$, $\varphi (u,u_{1,0})$, $\varphi (u_{1,0},u_{2,0})$, $\dots$, $\varphi (u_{m-1,0},u_{m,0})$:
\[ g= h(u, \varphi (u,u_{1,0}), \varphi (u_{1,0},u_{2,0}), \dots, \varphi (u_{m-1,0},u_{m,0})).\]
It is proved above that $X(g)=0$. Taking this fact into account, we obtain $h_u=0$ because $X(g)=X(h)=T_j^{-1}[ h_u (u_{0,1}, \psi (u,u_{1,0}), \dots, \psi (u_{m-1,0},u_{m,0}))] = h_u$. This means that the transformation~\eqref{itr} is equivalent to the composition of the transformation $w=\varphi (u,u_{1,0})$ and the transformation $v=h(w, w_{1,0}, w_{2,0}, \dots, w_{m-1,0})$. The latter transformation is invertible because
\[w=\varphi (P_0,P_1) = \tilde{P}_0 \big(g, T_j(g), \dots, T_j^{\tilde{d}}(g)\big)=\tilde{P}_0 \big(h, T_j(h), \dots, T_j^{\tilde{d}}(h)\big)\]
by virtue of equation~\eqref{pl}. The expressions for other dynamical variables can be obtained by the formulas $w_{0,r}=T_j^r(\tilde{P}_0)$ and $w_{s,0}=T_i^s(\tilde{P}_0)$.

Let $\varphi$ and $\psi$ be functionally dependent. Under this assumption equation~\eqref{abd} has the form $\varphi (u_{0,1},u_{1,1}) = E(\varphi (u_,u_{1,0}))$ and all functions $T_j^r(g)$ can be expressed in terms of $\varphi (u,u_{1,0})$, $\varphi (u_{1,0},u_{2,0})$, $\dots$, $\varphi (u_{m-1,0},u_{m,0})$ ($T_j(g) = h (E(\varphi (u,u_{1,0})), \dots, E(\varphi (u_{m-1,0},u_{m,0}))$ and so on). Hence the set $\{g, T_j(g), \dots, T_j^{\tilde{d}}(g)\}$ contains no more than $m$ functionally independent functions. But we prove above that this set must contain $m+1$ functionally independent functions if the transformation is invertible. Therefore, $\varphi$ and $\psi$ must be functionally independent if equation~\eqref{abd} admits an invertible transformation of the form $v=f(u, u_{1,0}, u_{2,0}, \dots, u_{m,0})$. \end{proof}

It is not always easy to see whether equation~\eqref{uij} can be represented in the form~\eqref{abd}. For example, at f\/irst glance it seems that the equation
\begin{gather}\label{tNY}
v_{1,1}=\frac{v (v_{1,0} + 1)}{v (v_{0,1}-v_{1,0}) + v_{0,1} + 1}
\end{gather}
does not admit an invertible transformation of the form $u=\varphi (v,v_{1,0})$. But in reality we can rewrite this equation as
\[\frac{v_{1,1} + 1}{v_{0,1} v_{1,1} - 1} = \frac{v+1}{v_{1,0} v -1}\]
and relate it to the equation
\begin{gather}\label{NY}
(u_{1,1} -1) (u_{0,1} + 1) = (u_{1,0} +1) (u - 1)
\end{gather}
via the invertible transformation
\[u= - 2 \frac{v_{1,0} + 1}{v v_{1,0} - 1} - 1,\qquad v=\frac {u_{0,1} - 1}{u + 1} .\]
Therefore, it is useful to reformulate our result in the following form.
\begin{corollary}\label{cor}
The equation~\eqref{uij} admits a non-point invertible transformation of the form~\eqref{itr} into an equation of the form~\eqref{vij} if and only if both the conditions
\[\left(\frac{F_{u_{1,0}}}{F_u}\right)_{u_{0,1}}=0, \qquad F_u+ F_{u_{1,0}} T_j^{-1}(F_{u_{0,1}}) \ne 0\]
are satisfied.
\end{corollary}

\begin{proof} If equation~\eqref{uij} is represented in the form \eqref{abd}, then the right-hand side $F$ of \eqref{uij} is determined as an implicit function from the identity
\begin{gather}\label{fip}
\varphi (u_{0,1},F)=\psi (u,u_{1,0}).
\end{gather}
Dif\/ferentiating this identity with respect to $u$ and $u_{1,0}$, we obtain
\begin{gather}\label{fip2}
T_j(\varphi _{u_{1,0}} (u,u_{1,0})) F_u=\psi _u (u,u_{1,0}), \qquad T_j(\varphi _{u_{1,0}} (u,u_{1,0})) F_{u_{1,0}} =\psi _{u_{1,0}} (u,u_{1,0}).
\end{gather}
Therefore, $F_{u_{1,0}}/F_u$ does not depend on $u_{0,1}$. Conversely, if $F_{u_{1,0}}/F_u = \beta(u,u_{1,0})$, then $F=E(\psi (u,u_{1,0}), u_{0,1})$ and \eqref{uij} can be rewritten in the form \eqref{abd}.

Dif\/ferentiating \eqref{fip} with respect to $u_{0,1}$, we obtain $F_{u_{0,1}}= -T_j\left(\varphi _u /\varphi _{u_{1,0}}\right)$. This expression and equation~\eqref{fip2} allow us to rewrite the functional independence condition for $\varphi$, $\psi$ in the following way
\begin{gather*}
\varphi _{u_{1,0}} (u,u_{1,0}) \psi _u (u,u_{1,0}) - \varphi _u (u,u_{1,0}) \psi _{u_{1,0}} (u,u_{1,0})  \\
\qquad {} = \varphi _{u_{1,0}} ( \psi _u + T_j^{-1}(F_{u_{0,1}}) \psi _{u_{1,0}})
= T_j(\varphi _{u_{1,0}}) \varphi _{u_{1,0}} ( F_u + T_j^{-1}(F_{u_{0,1}}) F_{u_{1,0}}) \ne 0. \tag*{\qed}
\end{gather*}
\renewcommand{\qed}{}
\end{proof}

Naturally, the propositions analogous to Theorem~\ref{form} and Corollary~\ref{cor} are valid for invertible transformations of the form $v=f(u, u_{0,1}, u_{0,2}, \dots, u_{0,n})$ too.

Returning to equations~\eqref{tNY}, \eqref{NY}, we note that equation~\eqref{NY} was introduced in \cite{NC} in a slightly dif\/ferent form. This equation has also been used in \cite{LY} as an example of an equation which is inconsistent around the cube (in the sense of \cite{ABS}) but possesses the higher symmetries. Therefore, we can obtain symmetries of equation~\eqref{tNY} from symmetries of equation~\eqref{NY}.

Indeed, if a transformation $v=f(u,u_{0,1})$ maps equation~\eqref{uij} into equation~\eqref{vij}, then dif\/ferentiation of \eqref{dtr} with respect to $\tau$ by virtue of a symmetry $u_{\tau}=\xi [u]$ of equation~\eqref{uij} gives us
\[ L_G\left( (f_{u_{0,1}} T_j + f_u )(\xi [u]) \right) = (\lambda [u] T_j + \mu [u]) \big( L_F (\xi [u]) \big),\]
where
\[ L_{G} = T_i T_j + G_{v_{1,0}} T_i + G_{v_{0,1}} T_j + G_v, \qquad  L_{F}=  T_i T_j + F_{u_{1,0}} T_i + F_{u_{0,1}} T_j + F_u.\]
Because $L_F(\xi [u])=0$ by def\/inition of symmetry, we see that $v_{\tau}= f_{u_{0,1}} T_j(\xi [u]) + f_u \xi [u]$ (after rewriting in terms of $v$ and its shifts) is a symmetry of equation~\eqref{vij}. Applying this, for example, to the three-point symmetries
\[ u_{\tau} =  (u^2 - 1) (u_{1,0} - u_{-1,0}), \qquad u_{\tau} = (u^2 - 1) \left( \frac{1}{u_{0,1} + u} - \frac{1}{u + u_{0,-1}} \right) \]
of equation~\eqref{NY}, we obtain the symmetries
\[ v_{\tau} = (v+1)^2 \left( \frac{1}{v v_{1,0} - 1} - \frac{1}{v v_{-1,0} - 1} \right), \qquad v_{\tau}= v \left( \frac{1}{v_{0,1} + 1} - \frac{1}{ v_{0,-1} + 1} \right)\]
of equation~\eqref{tNY}.

The Hietarinta \cite{hiet} equation\footnote{We write this equation in the form used in \cite{ram}.}
\begin{gather}\label{hiet}
u_{1,1}(u+\beta)(u_{0,1}+\alpha)=u_{0,1} (u+\alpha)(u_{1,0}+\beta)
\end{gather}
is another interesting example. The invertible transformations
\[v=\frac{u_{1,0} (u+\alpha)}{u} - \alpha,\qquad w=\frac{\beta u_{0,1}}{\beta+u-u_{0,1}}\]
map this equation into equation~\eqref{hiet} again. In addition, the Hietarinta equation is lineari\-zable~\cite{ram}. We note that the above properties of equation~\eqref{hiet} are similar to those of the continuous equation
\[u_{xy}=(\alpha(x,y) e^{u})_x + (\beta(x,y) e^{-u})_y + \gamma(x,y)\]
which was considered in \cite{SokSv}.

\section{Invertible transformations of dif\/ferential-dif\/ference equations}

We let $T$ denote the operator of the forward shift in $i$ by virtue of equation~\eqref{uix}. This operator is def\/ined by the following rules: $T(f(a,b,c,\dots))=f(T(a),T(b),T(c),\dots)$ for any function $f$;
$T(u_m)=u_{m+1}$; $T(u^{(n)})=D_x^{n-1}(F)$ (mixed variables $u_1^{(n)}$ are expressed in terms of the dynamical variables by virtue of equation~\eqref{uix}). Here
\[ D_x = \frac{\partial}{\partial x} + u^{(1)}\frac{\partial}{\partial u} + \sum_{k=1}^{\infty} \left( u^{(k+1)} \frac{\partial}{\partial u^{(k)}} + T^{(k-1)}(F) \frac{\partial}{\partial u_k} + T^{(1-k)}(\tilde{F}) \frac{\partial}{\partial u_{-k}}  \right),\]
i.e.\ $D_x$ is the total derivative with respect to $x$ by virtue of equations~\eqref{uix}, \eqref{umx}. The inverse (backward) shift operator $T^{-1}$ is def\/ined in the similar way.

\begin{definition}\label{sdtran}
We say that a transformation $v=f[u]$ maps equation~\eqref{uix} into an equation
\begin{gather}\label{vix}
{(v_1)}_x=G(x,v,v_1,v_x), \qquad G_{v_x} \ne 0
\end{gather}
if $D_x T (f)=G(x,f,T(f),D_x(f))$.
\end{definition}

\begin{definition}\label{sit}
A transformation $v=f[u]$ of equation~\eqref{uix} is called invertible if any of the dynamical variables $u$, $u_{k}$, $k \in \mathbb Z$, $u^{(l)}$, $l \in \mathbb N$ can be expressed as a function of a f\/inite subset of the variables
\[ x,\ \ v:=f,\  \ v_{r}:=T^{r}(f),\  \ v^{(s)}:=D_x^s(f),\  \ r \in \mathbb Z,\  \ s \in \mathbb N.\]
\end{definition}

\begin{definition}\label{sequiv}
Transformations $v=f[u]$ and $w=g[u]$ are called equivalent if there exists $r \in \mathbb Z$ such that $g= T^r(f)$.
\end{definition}

\begin{proposition}\label{sdchar}
Let a transformation of the form $v=g[u]$ be invertible and map equation~\eqref{uix} into equation~\eqref{vix}. Then this transformation is equivalent to either a transformation of the form
\begin{gather}\label{sditr}
w=f(x,u, u_1, u_2, \dots, u_m),
\end{gather}
or a transformations of the form
\begin{gather}\label{xtr}
w=f\big(x,u, u^{(1)}, u^{(2)}, \dots, u^{(n)}\big).
\end{gather}
\end{proposition}

\begin{definition}
A transformation is called non-point if this transformation is not equivalent to any point transformation of the form $w=g(x,u)$.
\end{definition}
It is easy to see that a transformation of the form \eqref{sditr} or \eqref{xtr} is non-point only if $f$ depends on more than one of the variables  $u, u_1, \dots,u_m$ or on at least one of the variables $u^{(1)}, \dots, u^{(n)}$, respectively.

\begin{theorem}\label{sdform}
Let a non-point invertible transformation of the form~\eqref{xtr}
map equation~\eqref{uix} into equation~\eqref{vix}. Then equation~\eqref{uix} can be written in the form $\varphi (x, u_1, {(u_1)}_x)=\psi (x, u, u_x)$, where $\varphi (x,y,z)$ and $\psi (x,y,z)$ satisfy the condition $\varphi _y \psi _z - \varphi _z \psi _y \ne 0$, and the transformation is equivalent to the composition of the invertible transformation $w=\varphi (x, u, u_x)$ and an invertible transformation of the form $v=h(x, w, w^{(1)}, w^{(2)}, \dots, w^{(n-1)})$. In particular, any non-point invertible transformation of the form $v=f(x, u, u_x)$ is equivalent to the composition of the transformation $w=\varphi (x, u, u_x)$ and a point transformation $v=h(x,w)$.
\end{theorem}

\begin{corollary}\label{xcor}
The equation~\eqref{uix} admits a non-point invertible transformation of the form~\eqref{xtr} into an equation of the form~\eqref{vix} if and only if both the conditions
\[ F_u F_{u_x u_1} - F_{u u_1} F_{u_x} = 0, \qquad F_u+ F_{u_x} T^{-1}(F_{u_1}) \ne 0\]
are satisfied.
\end{corollary}

\begin{theorem}\label{sdform1}
Let a non-point invertible transformation of the form~\eqref{sditr} map equation~\eqref{uix} into equation~\eqref{vix}. Then equation~\eqref{uix} can be written in the form $D_x( p(x,u,u_1))=q(x, u, u_1)$, where $p(x,y,z)$ and $q(x,y,z)$ satisfy the condition $p_y  q_z - p_z q_y \ne 0$, and the transformation is the composition of the transformation $w=p(x,u,u_1)$ and an invertible transformation of the form $v=h(x, w, w_1, w_2, \dots, w_{m-1})$. In particular, any non-point invertible transformation of the form $v=f(x, u, u_1)$ is the composition of the transformation $w=p(x, u, u_1)$ and a point transformation $v=h(x,w)$.
\end{theorem}

\begin{corollary}
The equation~\eqref{uix} admits a non-point invertible transformation of the form~\eqref{sditr} into an equation of the form~\eqref{vix} if and only if equation~\eqref{uix} has the form
\[\left(u_1\right)_x = a(x, u, u_1) u_x + b(x, u, u_1),\]
where $a$ and $b$ satisfy the condition $a_x + a_{u_1} b - a b_{u_1} - b_u \ne 0$.
\end{corollary}

For brevity, we omit the proofs of the above propositions because they are very similar to the proofs for discrete equations.

\section{Examples: the transformations\\ of Liouville equation analogues}\label{s4}
A special class of integrable equations of the form
\begin{gather}\label{uxy}
u_{xy}=F(x,y,u,u_x,u_y)
\end{gather}
consists of equations for which there exist both a dif\/ferential substitution of the form $v=X(x,y,u_x,u_{xx}, \dots)$ and a substitution of the form $w=Y(x,y,u_y,u_{yy}, \dots)$ that map \eqref{uxy} into the equations $v_y=0$ and $w_x=0$, respectively. Such equations are called Darboux integrable or equations of the Liouville type. They not only are C-integrable (in accordance with the term of \cite{Cal}) but also possess inf\/initely many symmetries of arbitrary high order \cite{ZSS,ZS}. The complete classif\/ication of the Darboux integrable equations \eqref{uxy} has been performed in \cite{ZS}. Equations with the analogous properties exist among equations of the form \eqref{uij} and \eqref{uix} too, but the classif\/ication of such equations is completed for a special case of equation~\eqref{uix} only \cite{HabZ}. Therefore, deriving new examples of discrete and semi-discrete Darboux integrable equations from already known equations may be useful (for example, to check the completeness of a future classif\/ication).

\subsection{Discrete equations}
The f\/irst example is the discrete Liouville equation
\begin{gather}\label{ld}
u_{1,1}=\frac{(u_{1,0}-1)(u_{0,1}-1)}{u}
\end{gather}
from \cite{Hirota}. According to \cite{AdS}, this equation has the integrals
\begin{gather}\label{IJ}
I[u]=\left( \frac{u_{2,0}}{u_{1,0}-1} +1 \right) \left( \frac{u-1}{u_{1,0}} + 1 \right),\qquad J[u]=\left( \frac{u_{0,2}}{u_{0,1}-1} +1 \right) \left( \frac{u-1}{u_{0,1}} + 1 \right),
\end{gather}
i.e.\ functions $I[u]$, $J[u]$ such that $T_j(I[u])=I[u]$, $T_i(J[u])=J[u]$. In addition, equation~\eqref{ld} is linearizable: the substitution
\begin{gather}\label{dwt}
u= \frac{z_{0,1} z_{1,0}}{(z_{1,0}-z)(z_{0,1}-z)}
\end{gather}
maps solutions of the equation
\begin{gather}\label{dwave}
z_{1,1}=z_{1,0}+z_{0,1}-z
\end{gather}
into solutions of \eqref{ld}.

It is easy to see that \eqref{ld} can be written in the form $T_j(\varphi (u,u_{1,0}))= \psi (u,u_{1,0})$ and the scheme \eqref{abd}--\eqref{pqd} is applicable to this equation:
\begin{gather*}
 v=\varphi =\frac{u_{1,0}}{u-1} ,\qquad v_{0,1}=\psi =\frac{u_{1,0}-1}{u},\\
 u=p=\frac{v+1}{v-v_{0,1}} ,\qquad u_{1,0}=q=v\frac{v_{0,1}+1}{v-v_{0,1}}, \\
 \frac{v_{1,0}+1}{v_{1,0}-v_{1,1}}= \frac{v_{0,1}+1}{v-v_{0,1}} v.
 \end{gather*}
Thus, we obtain the equation
\begin{gather}\label{tld}
v \frac{v_{1,1}- v_{1,0}}{v_{0,1}-v} = \frac{v_{1,0}+1}{v_{0,1}+1}
\end{gather}
that is related to the discrete Liouville equation via the invertible transformation $v=u_{1,0}/(u-1)$. Substituting the expressions of $u, u_{0,1}, u_{1,0}, \dots$ in terms of $v, v_{0,1}, v_{1,0}, \dots$ into \eqref{IJ}, we obtain the integrals of equation~\eqref{tld}:
\[
I[v]=v_{1,0} + \frac{v_{1,0}+1}{v}, \qquad J[v]=\frac{(v_{0,3}-v_{0,1})(v_{0,2}-v)}{(v_{0,3}-v_{0,2})(v_{0,1}-v)}.
\]
The composition
\[v=\frac{z_{0,2} (z_{1,0}-z)}{z (z_{2,0}-z_{1,0})}\]
of the transformation $v=u_{1,0}/(u-1)$ and \eqref{dwt} allows us to construct the solution
\[v=\frac{(\alpha _{i+2} +\beta _j)(\alpha _{i+1}-\alpha _i )}
{(\alpha _i +\beta _j) (\alpha _{j+2} - \alpha _{i+1})}\]
of equation~\eqref{tld} from the general solution $z= \alpha _i + \beta _j$ of \eqref{dwave}, where $\alpha _i$ and $\beta _j$ are arbitrary.

Equation~\eqref{tld} can be written in the form $T_j(\varphi)=\psi$ but $\varphi$ and $\psi$ are functionally dependent ($\varphi =\psi = I[v]$). According to Theorem~\ref{form}, this fact implies that \eqref{tld} has no non-point invertible transformation of the form $\tilde v = f(v,v_{1,0},\dots,v_{n,0})$ and hence  equation~\eqref{ld} admits, up to equivalence, only the f\/irst order invertible transformations ($v=f(u_{1,0}/(u-1))$ and $w=g(u_{0,1}/(u-1))$ only).

Applying Corollary~\ref{cor}, we see that the other discrete version \cite{Hir79} of the Liouville equation
\[ v_{1,1}=\frac{v_{1,0}v_{0,1} - 1}{v}\]
does not admit a non-point invertible transformation. This equation is mapped into \eqref{ld} via the non-invertible transformation $u=v_{1,0}v_{0,1}$ and has the integrals
\[ I[v]=\left(\frac{v_{3,0}}{v_{1,0}} +1 \right)\left(\frac{v}{v_{2,0}} +1 \right),\qquad J[v]=\left(\frac{v_{0,3}}{v_{0,1}} +1 \right)\left(\frac{v}{v_{0,2}} +1 \right).\]

\subsection{Dif\/ferential-dif\/ference equations}

Let us consider the following analogue of the Liouville equation:
\begin{gather}\label{sdl}
{(u_1)}_x = u_1 \left(u_1 + \frac{u_x}{u} + u\right).
\end{gather}
This equation has the integrals
\[X[u]= 2 \frac{u_{xx}}{u} - 3 \frac{u_x^2}{u^2} - u^2,\qquad I[u]=\left(1 + \frac{u_1}{u_2}\right) \left(1+\frac{u_1}{u}\right),\]
i.e.\ functions $X[u]$, $I[u]$ such that $T(X)=X$, $D_x(I)=0$. Like the discrete and continuous Liouville equations, equation~\eqref{sdl} is linearizable: the substitution
\begin{gather}\label{sdwt}
u= \frac{(z_1 - z) z_x}{z_1 z}
\end{gather}
maps solutions of the equation
\begin{gather}\label{sdwave}
{(z_1)}_x =z_x
\end{gather}
into solutions of \eqref{sdl}. The above information and some other details about equation~\eqref{sdl} can be found in \cite{AdS}.

Equation~\eqref{sdl} can be written as
\[ \frac{{(u_1)}_x}{u_1} - u_1 = \frac{u_x}{u} + u.\]
Applying the scheme~\eqref{ab}--\eqref{pq}, we obtain
\begin{gather}
 v=\frac{1}{2} \left(\frac{u_x}{u} - u\right),\qquad v_1 = \frac{1}{2} \left(\frac{u_x}{u} + u\right),\nonumber\\
 u=v_1 - v,\qquad u_x = v_1^2-v^2,\nonumber\\
\label{ricc}
{(v_1 - v)}_x = v_1^2-v^2.
\end{gather}

Thus, the invertible transformation $v= (u_x/u - u)/2$ maps equation~\eqref{sdl} into the sequence of the coupled Riccati equations~\eqref{ricc}. Expressing $X[u]$ and $I[u]$ in  terms of $v, v_1, v_x, \dots$, we obtain the integrals
\[ X[v]=v_x - v^2,\qquad I[v]=\frac{(v_3-v_1)(v_2-v)}{(v_3-v_2)(v_1-v)}.\]
of equation~\eqref{ricc}. The composition
\[v=\frac{z_{xx}}{2 z_x} - \frac{z_x}{z}\]
of the invertible transformation and \eqref{sdwt} generates the solution
\begin{gather}\label{ricsol}
v= \frac{\beta_{xx}}{2 \beta_x} - \frac{\beta_x}{\alpha_i + \beta}
\end{gather}
of equation~\eqref{ricc} from the general solution $z= \alpha_i + \beta (x)$ of \eqref{sdwave}, where $\alpha _i$ and $\beta (x)$ are arbitrary. Equation~\eqref{ricc} was used in \cite{AdS} as an example of an equation admitting the integrals and the solution~\eqref{ricsol} was constructed in this article by another method (directly form the equation $X[v]=\xi (x)$).

Moreover, equation~\eqref{sdl} can be represented in the form~\eqref{pq} too. Applying the scheme \eqref{ab}--\eqref{pq} in the reverse order, we get
\begin{gather*}
w=p=\frac{u_1}{u} ,\qquad w_x=q=\frac{u_1^2}{u} + u_1,\\
 u=\frac{w_x}{(w+1)w},\qquad u_1=\frac{w_x}{w+1},\\
 \frac{(w_1)_x}{(w_1+1)w_1} = \frac{w_x}{w+1}
 \end{gather*}
and see that the invertible transformation $w= u_1 / u$ maps \eqref{sdl} into the equation
\begin{gather}\label{tsdl}
(w_1)_x = w_x w_1 \frac{w_1+1}{w+1}\,.
\end{gather}
As above, we construct the integrals
\[
X[w]=2 \frac{w_{xxx}}{w_x} - 3 \frac{w_{xx}^2}{w_x^2},\qquad I[w]=\frac{(w_1+1)(w+1)}{w_1}
\]
of equation~\eqref{tsdl} by expressing $X[u]$ and $I[u]$ in terms of $w, w_1, w_x, \dots$, and obtain its solution
\[w= \frac{(\alpha _{i+2} - \alpha _{i+1}) (\alpha _i + \beta (x))}{(\alpha _{i+1} - \alpha _i) (\alpha _{i+2} + \beta (x))}\]
with arbitrary $\alpha _i$ and $\beta (x)$  by applying the composition
\[w=\frac{(z_2 - z_1) z}{(z_1 - z) z_2}\]
of the transformations $w= u_1 / u$ and \eqref{sdwt} to the general solution $z=\alpha _i +\beta (x)$ of equation~\eqref{sdwave}.

The semi-discrete Liouville equation~\eqref{sdl} is a special case of the Darboux integrable equation
\begin{gather}\label{hab}
\left(u_1\right) _x = u_x + \sqrt{C e^{2u_1} + B e^{(u_1+u)} + C e^{2u}} .
\end{gather}
that was introduced in \cite{HabZ}. Indeed, replacing $u$ in \eqref{sdl} by $\exp (u)$, we obtain equation~\eqref{hab} with $C=1$, $B=2$. Without loss of generality, we can assume that the constant $C$ in equation~\eqref{hab} equals $1$ or $0$ because $C$ can be scaled via the point transformation $u=\tilde u + \gamma$. Applying Corollary~\ref{xcor}, we see that equation~\eqref{hab} admits an invertible transformation of the form $v=f(x,u,u_x)$ only if $B=2C$. The invertible transformations
\[ w= e^{u_1 - u},\qquad e^u = \frac{w_x}{w \sqrt{C w^2 + B w + C}}\]
relate \eqref{hab} to the equation
\begin{gather}\label{mw}
\left( w_1\right)_x = w_1 w_x \sqrt{\frac{C w^2_1 + B w_1 + C}{C w^2 + B w + C}}\,.
\end{gather}
The later equation has the integrals
\begin{gather*}
X[w]=2\frac{w_{xxx}}{w_x} - 3 \left(\frac{w_{xx}}{w_x}\right) ^2 + \frac{3 w_x^2 (B^2 - 4C^2)}{4 (C w^2 + B w + C)} ,\\
  I[w]= \int^{w_1} \frac{ds}{s\sqrt{C s^2 + B s + C}} - \int^{w} \frac{ds}{\sqrt{C s^2 + B s + C}}
  \end{gather*}
and can not be reduced to equation~\eqref{tsdl} via a point transformation because equation~\eqref{mw}, in contrast to equation~\eqref{tsdl}, does not admit an invertible transformation of the form $v=f(x,w,w_x,w_{xx})$ if $B \ne 2 C$.

\subsection*{Acknowledgments}

This work is partially supported by the Russian Foundation for Basic Research (grant number 10-01-00088-a).

\pdfbookmark[1]{References}{ref}
\LastPageEnding

\end{document}